%% file: SAIP2021.tex
\documentclass[a4paper]{jpconf}
\usepackage{graphicx}
\usepackage{caption}
\usepackage{subcaption}
\begin{document}
\title{An investigation of over-training within semi-supervised machine learning models in the search for heavy resonances at the LHC}

\author{Benjamin Lieberman$^1$, Joshua Choma$^1$, Salah-Eddine Dahbi$^1$, Bruce Mellado$^{1,2}$ and Xifeng Ruan$^1$}

\address{$^1$School of Physics and Institute for Collider Particle Physics, University of the
Witwatersrand, Johannesburg, Wits 2050, South Africa}
\address{$^2$iThemba LABS, National Research Foundation, PO Box 722, Somerset West 7129,
South Africa}

\ead{benjamin.lieberman@cern.ch}

\begin{abstract}
In particle physics, semi-supervised machine learning is an attractive option to reduce model dependencies searches beyond the Standard Model. When utilizing semi-supervised techniques in training machine learning models in the search for bosons at the Large Hadron Collider, the over-training of the model must be investigated. Internal fluctuations of the phase space and bias in training can cause semi-supervised models to label false signals within the phase space due to over-fitting. The issue of false signal generation in semi-supervised models has not been fully analyzed and therefore utilizing a toy Monte Carlo model, the probability of such situations occurring must be quantified. This investigation of $Z\gamma$ resonances is performed using a pure background Monte Carlo sample. Through unique pure background samples extracted to mimic ATLAS data in a background-plus-signal region, multiple runs enable the probability of these fake signals occurring due to over-training to be thoroughly investigated.
\end{abstract}


\input{./Introduction}


\input{./Methodology}


\input{./Results}

\input{./Conclusions}

\section*{References}
\bibliographystyle{iopart-num}
\bibliography{bibliog}

\end{document}

%% file: Introduction.tex
\section{Introduction}
In 2012 the ATLAS and CMS collaborations reported on the observation of a Higgs Boson with an mass of 125\,GeV~\cite{ATLAS:2012yve,CMS:2012qbp}. This discovery further motivates the search for new bosons. 

A 2HDM+$S$ model, where $S$ is a singlet scalar, was used in Ref.~\cite{vonBuddenbrock:2015ema,vonBuddenbrock:2016rmr} to explain some features of the Run 1 Large Hadron Collider (LHC) data. Here the heavy scalar, $H$, decays predominantly into $SS,Sh$, where $h$ is the SM Higgs boson. The model predicts the emergence of multi-lepton anomalies that have been verified in Refs.~\cite{vonBuddenbrock:2017gvy,Buddenbrock:2019tua,vonBuddenbrock:2020ter,Hernandez:2019geu}, where a possible candidate of $S$ has been reported in Ref.~\cite{Crivellin:2021ubm}. The model can further elaborate on multiple anomalies in astro-physics if it is complemented by a candidate of a Dark matter~\cite{Beck:2021xsv}. It can be easily extended~\cite{Sabatta:2019nfg} to account for the $4.2\sigma$ anomaly $g-2$ of the muon~\cite{Muong-2:2021ojo,Aoyama:2020ynm} (see Ref.~\cite{Fischer:2021sqw} for a review of anomalies).
 
 The above mentioned motivates for the searches of heavy scalar resonances. We choose to investigate the search of $H\rightarrow Z\gamma$ with $Z\rightarrow\ell\ell$ and $\ell=e,\mu$. This is done using semi-supervision with topological features, as suggested in Ref.~\cite{Dahbi:2020zjw}. In this presentation we focus on the potential over-training entailed in the the use of semi-supervision when confronting side-bands and the signal region using a Deep Neural Network.  


\subsection{$Z\gamma$ Dataset}
\label{zgamma_dataset}

In the search for new bosons, the $Z\gamma$ final state data is used as a pure background sample. This is done as $Z\gamma$ represents almost $90\%$ of the total background. This is an ideal dataset to evaluate the extent of false signals generated during the Machine Learning training as any signals found within the dataset are a product of over-training and/or fluctuations within the phase space. 
The $Z\gamma$ dataset is described in detail in the ATLAS conference note, Ref.~\cite{ATLAS:2019nxl}. The final $Z\gamma$ variables selected for the analysis are as follows: invariant mass, $m_{\ell\ell\gamma}$;  invariant di-jet mass, $m_{jj}$; pseudo-rapidity of leading and sub-leading jets, $\eta_{j1}$, $\eta_{j2}$; number of jets, $N_{j}$; number of leptons, $N_{\ell}$; number of $b$-jets, $N_{bj}$ and missing transverse energy. The invariant mass $m_{\ell\ell\gamma}$ is used to select the mass range of the analysis.

\subsection{Machine Learning in Discovering Physics Beyond the Standard Model}

Machine Learning (ML) is a computational system that utilises algorithmic and statistical models in order to perform a specific task. It does this by finding patterns and anomalies within a dataset. In the case of the data produced at ATLAS, ML is an ideal tool to process the data, extracting the important information from the data so that it can be used for further analysis. It is therefore able to extract specific final state debris from a conglomerate of interacting systems that occurred at the particle collision.



The quantification of uncertainties propagated within ML methods is vital in sub-atomic physics analysis as it allows both an understanding of the accuracy of any predictions made and exposes the level of validity of any ML based discoveries. The uncertainties in fully supervised techniques used in particle physics are well defined and extensively researched, however the uncertainty propagated in semi-supervised techniques have not been quantified to the same extent. This research therefore focuses on measuring the uncertainties or fake signals produced in the training of semi-supervised models within a given phase space. 



\subsubsection{Semi-supervised Machine Learning}
Semi-supervised learning utilises aspects of both supervised and unsupervised learning styles. Semi-supervised learning techniques make use of a combination of labelled and unlabelled data during training.
The algorithm is therefore trained with a well known sample of labelled data together with an unlabelled sample. The algorithm can therefore learn to accurately discern events that match the labelled data as well as find patterns within the unlabelled events which aren't restricted by prior expectations. This takes advantage of both the successes of supervised and unsupervised learning~\cite{Dahbi:2020zjw}.

In particle physics, the labelled sample used is pure background and the unlabelled sample is comprised of a mix of signal and background processes. This allows signal processes and interactions to be extracted without having to conform to a preconceived definition of the signal. This in turn provides the potential of experimentally finding discrepancies between current theoretical expected results and those experimentally determined. This may therefore lead to an increased chance of understanding new particles.

%% file: Methodology.tex
\section{Methodology}

Through a comprehensive analysis, a Deep Neural Network (DNN) model is selected as the optimum classifier. The DNN is designed and optimised for both fully-supervised and semi-supervised applications. The quantification of uncertainties produced in semi-supervised ML techniques is fundamental in assessing the extent of validity of any semi-supervised sub-atomic particle physics result. 

A benchmark centre of mass of $200$\,GeV is selected and the data samples are divided into a mass-window ($194-206$\,GeV) and side-band ($194-182$\,GeV and $206-218$\,GeV) samples using the invariant mass.

\subsection{Deep Neural Network Model}
The DNN used in this experiment is optimised through validation of binary cross-entropy and accuracy in the training of the model as well as the area under the Receiver Operating Characteristic (ROC) curve, describing the classifier performance, output by the model. The model itself is kept simple initially and can later be adjusted to a more complex structures when further evaluation is required. 

A learning rate of $1 \cdot 10^{-3}$ is used with a learning decay of $3\cdot10^{-4}$. The model is run for $8$ epochs using a batch size of $1$. The initial optimised model is summarised in Table \ref{tab:dnn_model}. As the model is run using the $Z\gamma$ background, there is no significant separation between the mass-window and side-band samples. 

\begin{table}[t]
    \caption{Initial optimised DNN model.}
    \label{tab:dnn_model}
    \begin{center}
    \begin{tabular}{lll}
    \br
    Layer & Number of nodes  & Activation function\\
    \mr
    Input layer & 360 & Relu \\
    Hidden layer 1 & 180 & Relu \\
    Hidden layer 2 & 180 & Relu \\
    Hidden layer 3 & 90 & Relu \\
    Hidden layer 4 & 180 & Relu \\
    Output layer & 1 & Sigmoid \\
    \br
    \end{tabular}
    \end{center}
\end{table}

\subsection{Sample Generation}
\subsubsection{Toy Monte Carlo Model}

The Toy Monte Carlo (MC) sample generator is constructed to take the pre-processed $Z\gamma$ data and output a sample with accurate statistics mimicking data samples from the ATLAS experiment. Each toy MC sample generated uses randomly selected events from the input data to produce unique samples. In order for the sample to accurately mimic ATLAS data, the MC event weighting of each event is used. 

\subsection{Evaluating Over-training on Invariant Mass}
In order to calculate the significance of false signals being generated, the following method is applied to the response distribution output from the DNN.

\begin{enumerate}
    \item Event statistics are extracted from the response distribution to form the batches representing 50, 60, 70, 80 and 90\% of the background events.
    \item The invariant mass, $m_{\ell\ell\gamma}$, distribution of each batch is than analysed in terms of the mass-window and side-band. This is done by fitting an exponential function, Equation \ref{exponential_function}, and a exponential + Gaussian function, Equation \ref{exponential_Gaussian_function}, to each batch's invariant mass distribution:
    \begin{equation}
        f(x) = n_0 \cdot e^{ax + bx^2},
        \label{exponential_function}
    \end{equation}
    \begin{equation}
        g(x) = n_0 \cdot e^{ax + bx^2} + n_1 \cdot e^{\frac{(x - \mu)^2}{2\sigma}},
        \label{exponential_Gaussian_function}
    \end{equation}
    where $n_0$, $a$, $b$ and $n_1$ are constants produced in the fit; $\mu$ is the mean and $\sigma$ are the standard deviation.
    
\end{enumerate}

\subsection{Calculating Significance}
The significance of fake signals generated due to over training in the mass-window can than be quantified as the difference between the log-likelihoods of the two functions. The following steps are implemented:

\begin{enumerate}
    \item The log-likelihood can be calculated using a Poisson probability mass function, $p_X$, on the first $n$ terms of the random Poisson variables $\{X_n\}$. The probability mass function of a term $X_i$ is:
    \begin{equation}
        p_X(x_i) = e^{-\lambda_0 \frac{\lambda_0^{x_i}}{x_i!}},
        \label{poisson_prob_mass_funct}
    \end{equation}
    where $\lambda_0$ is the parameter of interest. The likelihood function, $L$, and log-likelihood, $\ln(L)$, can therefore be calculated as follows:
    \begin{equation}
        L(\lambda ; x_1, x_2, ... , x_n) = \prod_{i=1}^{n} e^{-\lambda} \frac{\lambda^{x_i}}{x_i!},
        \label{poisson_likelihood}
    \end{equation}
    \begin{equation}
        \ln L(\lambda ; x_1, x_2, ... , x_n) = -n\lambda - \sum_{i=1}^{n} \ln(x_i!) + \ln(\lambda) \sum_{i=1}^{n}x_i.
        \label{poisson_log_likelihood}
    \end{equation}

    \item The log-likelihood of the two functions can than be used to calculate the run's uncertainty significance:
    \begin{equation}
        S_k = \sqrt{2 \cdot (\ln L_{eg} - \ln L_{e})}, 
    \end{equation}
    where $S_k$ is the Significance for the $k^{th}$ run and $L_{eg}$ and $L_{e}$ are the log-likelihoods of the exponential + Gaussian function and the Exponential function, respectively.
    
    \item Repeating the process with statistically random toy MC samples a number of times (initially 500 times) will produce the statistical deviations in significance of fake signals being generated. The uncertainty generated, within the semi-supervised model, can therefore be quantify. As the samples are limited by the MC statistics, the number of runs is limited to 500.  
\end{enumerate}

%% file: Results.tex
\section{Results}

\subsection{Invariant Mass Distribution with Cuts}
\begin{figure}[t]
\centering
    \includegraphics[width=0.50\linewidth]{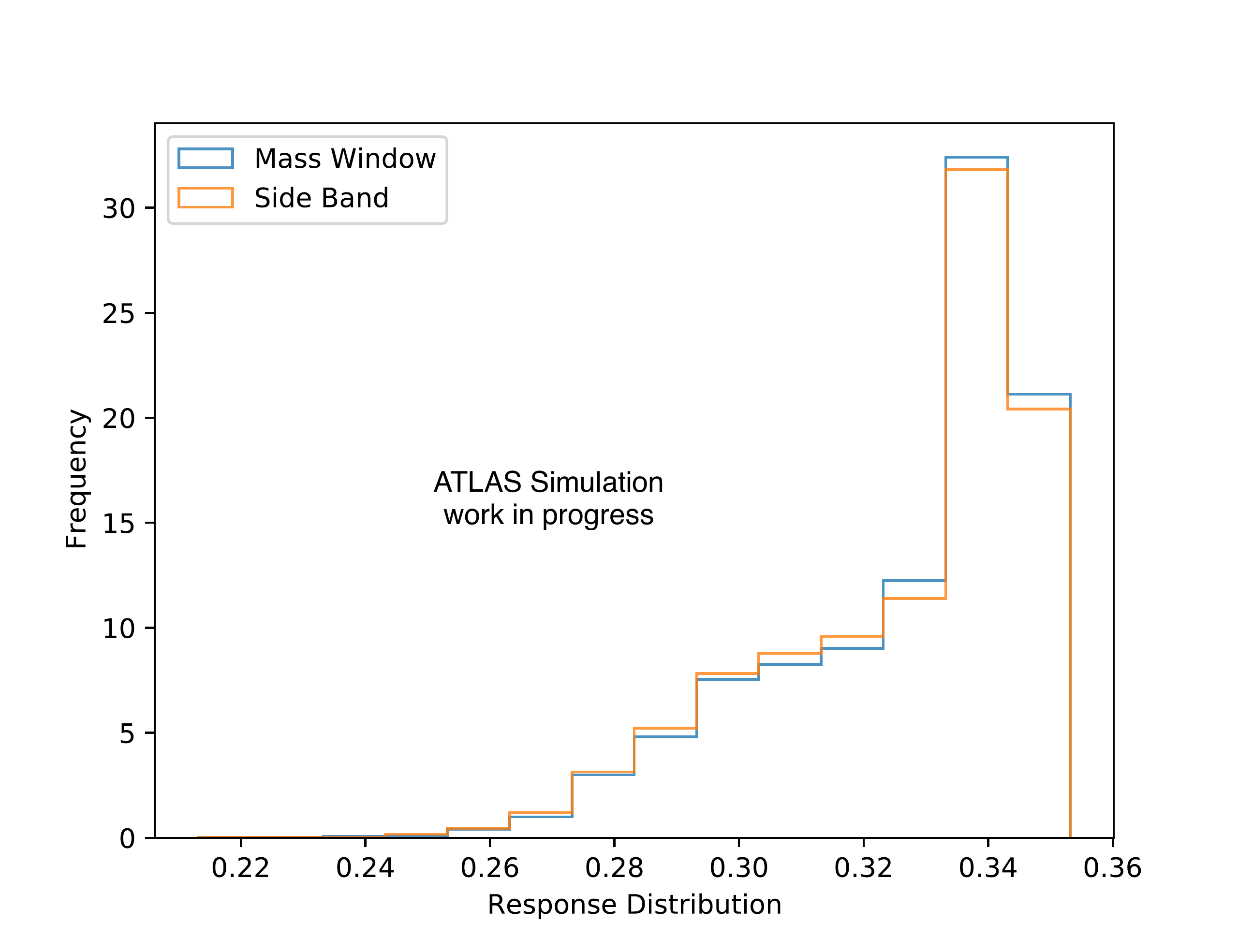}
    \caption{Example Response Distribution Output from DNN Toy Monte Carlo Model.}
    \label{fig:response_dist_zgamma}
\end{figure}

In order to analyse false signals generated in the training of the model, the output response distribution of the DNN is analysed, example in Figure \ref{fig:response_dist_zgamma}. 

\begin{figure}[t]
     \centering
     \begin{subfigure}[b]{0.45\textwidth}
         \centering
         \includegraphics[width=\linewidth]{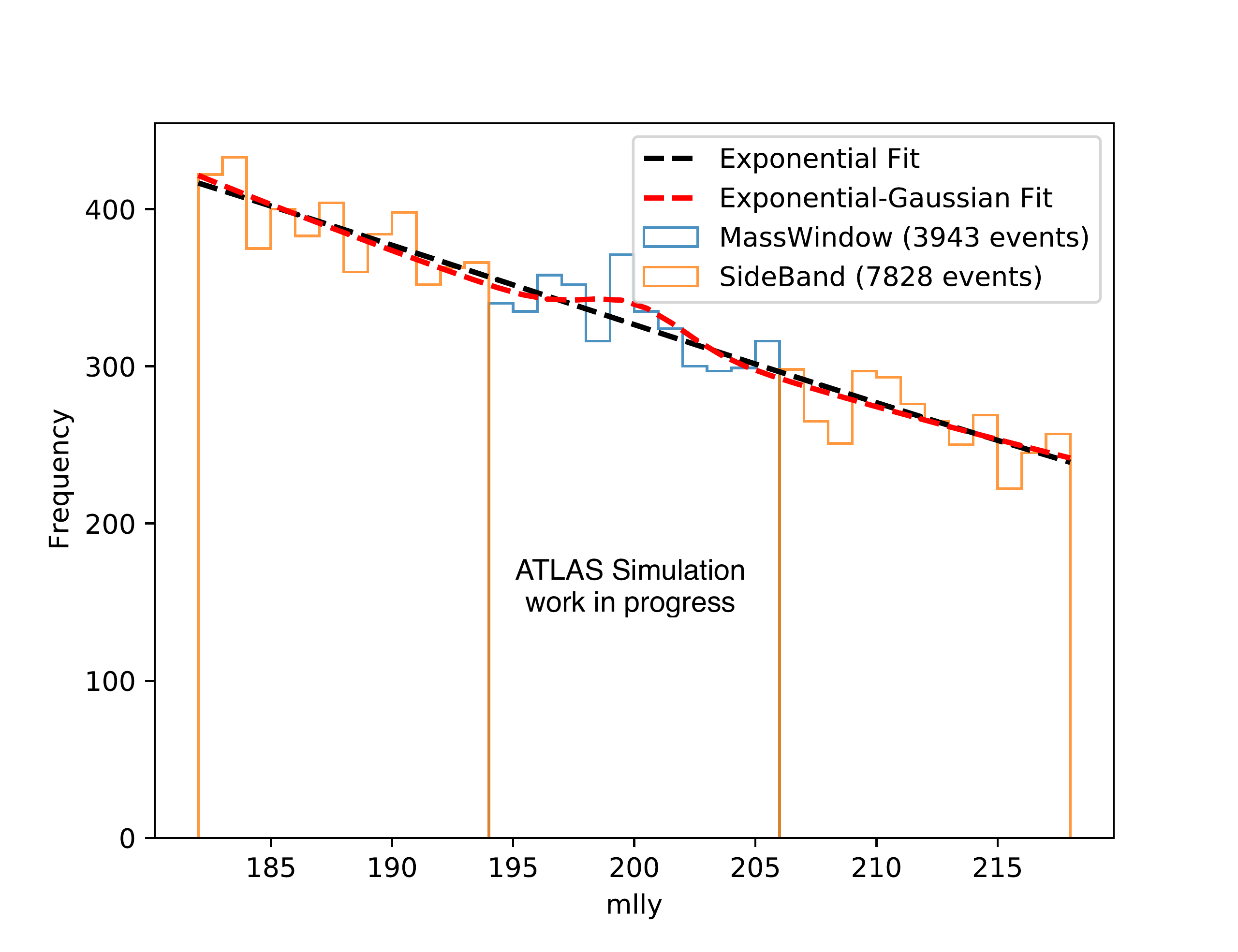}
         \label{fig:mlly_60}
     \end{subfigure}
     \hfill
     \begin{subfigure}[b]{0.48\textwidth}
         \centering
         \includegraphics[width=\linewidth]{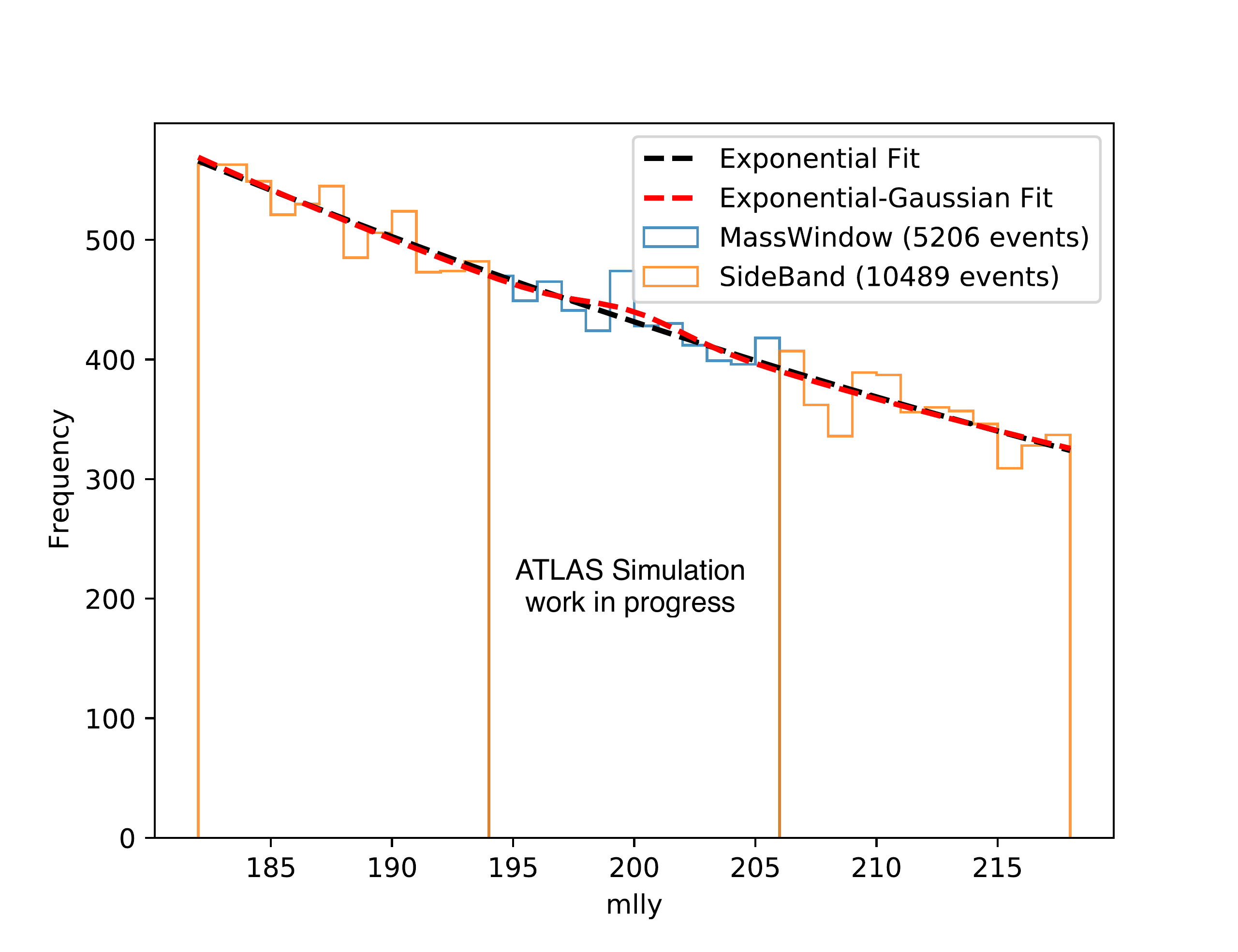}
         \label{fig:mlly_80}
     \end{subfigure}
        \caption{Example distributions of $m_{\ell\ell\gamma}$ for 60\% and 80\% background rejection.}
        \label{fig:mlly_dists}
\end{figure}

The response distribution is divided into batches containing $50$, $60$, $70$, $80$ and $90\%$ of the background events. Batches are filled starting from the response distribution's maximum, $1$, and move towards the minimum, $0$, until the required percentage of events are captured. Each event in the response distribution is mapped to it's corresponding invariant mass, and the fit functions are applied to the given distribution. The invariant mass distributions for the given example run are shown in Figure~\ref{fig:mlly_dists}.

\subsection{Significance Distributions}
For the given example run, the significance calculated using the difference in the log-likelihoods of the fit functions is summarised in Table \ref{tab:mlly_sig_summary}.

\begin{table}[h]
    \caption{Summary of $m_{\ell\ell\gamma}$ cut significance example.}
    \label{tab:mlly_sig_summary}
    \begin{center}
    \begin{tabular}{llll}
    \br
    \% Events & Mass-window events & Side-band events  & Significance ($\sigma$)\\
    \mr
    50 & 3282 & 6527 & 1.07 \\
    60 & 3943 & 7828 & 1.53\\
    70 & 4585 & 9148 & 1.07\\
    80 & 5206 & 10489 & 0.83\\
    90 & 5835 & 11822 & 0.58\\
    \br
    \end{tabular}
    \end{center}
\end{table}

\begin{figure}[t]
     \centering
     \begin{subfigure}[b]{0.47\textwidth}
         \centering
         \includegraphics[width=\linewidth]{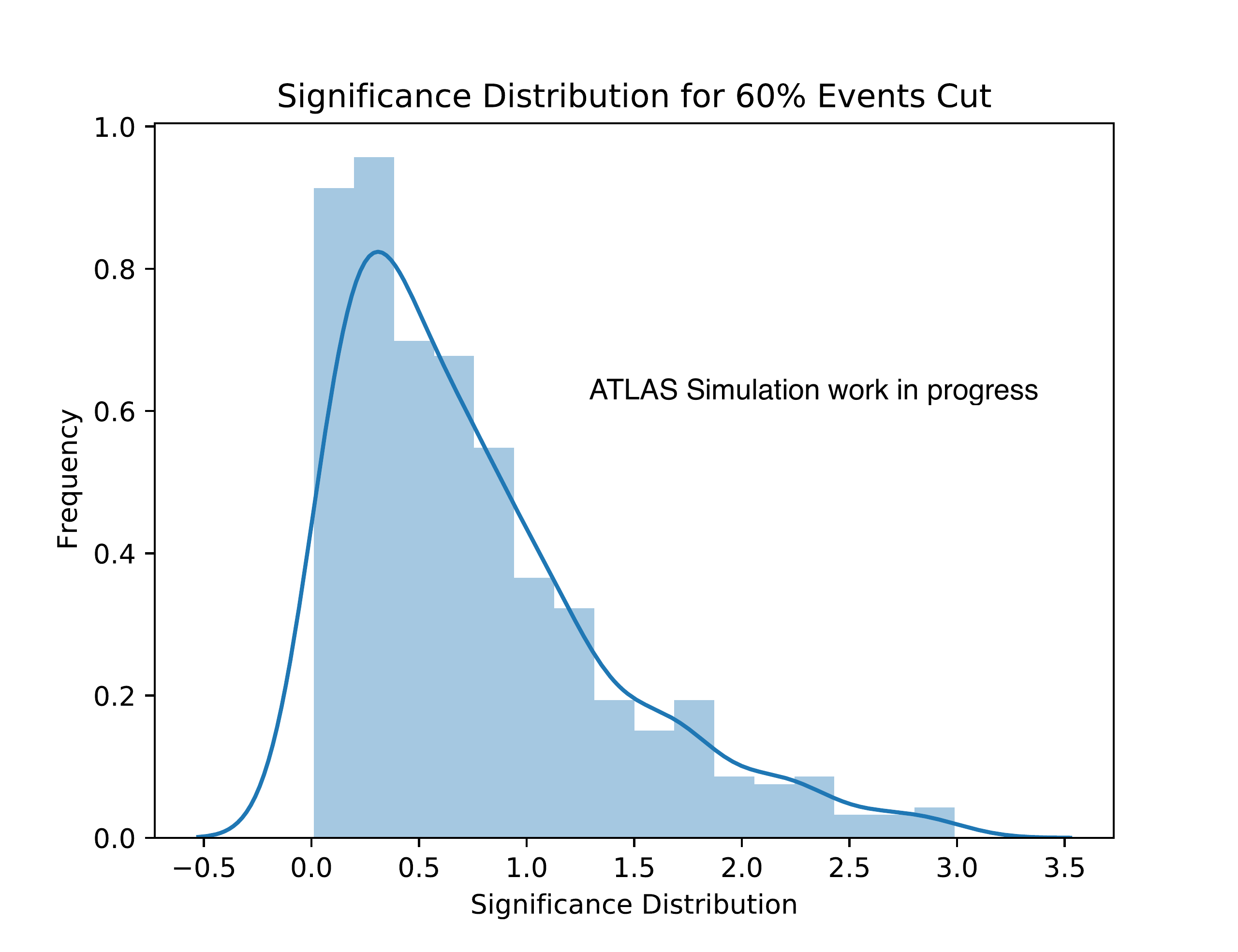}
         \label{fig:sig_60}
     \end{subfigure}
     \hfill
     \begin{subfigure}[b]{0.47\textwidth}
         \centering
         \includegraphics[width=\linewidth]{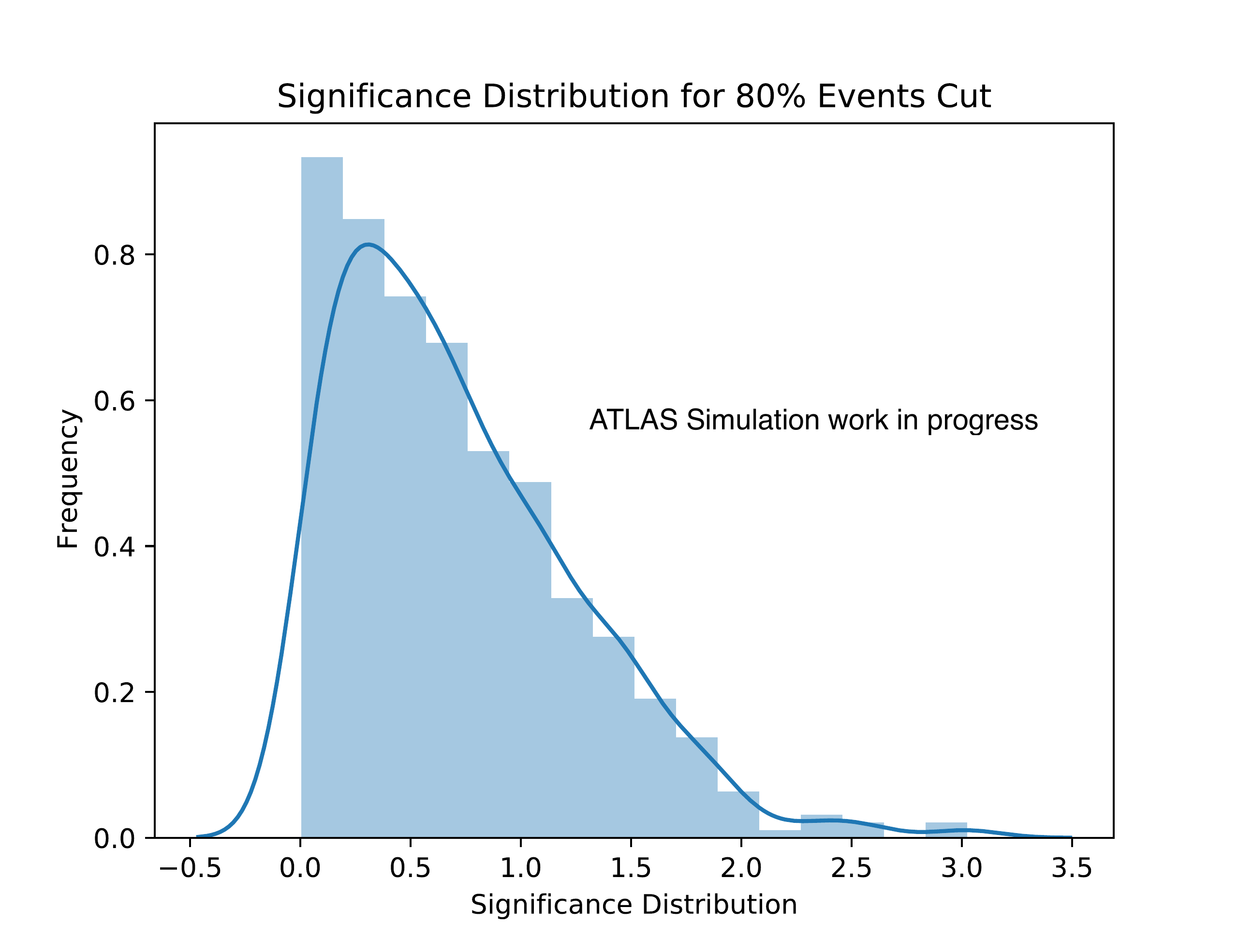}
         \label{fig:sig_80}
     \end{subfigure}
    \caption{Significance distribution on 60\% and 80\% background rejection for 500 runs.}
    \label{fig:sig_dists}
\end{figure}

Running the model multiple times on unique toy Monte Carlo samples produces distributions on the significance which can therefore be used to quantify the extent of false signals produced in the model. The results below, in Figure~\ref{fig:sig_dists}, demonstrate examples of the significance distributions produced when the model is run on 500 toy MC generated samples.

%% file: Conclusions.tex
\section{Conclusion}
The investigation into quantifying the uncertainty generated, through the over-training of semi-supervised techniques, using $Z\gamma$ resonances was performed using pure background toy MC generated samples and a semi-supervised DNN model. The invariant mass distributions for various background rejections was used to measure the fake signals produced by the model. This in turn was quantified through the calculated significance for each background rejection of each run. The significance distributions produced on 500 samples, Figure \ref{fig:sig_dists}, form the positive side of a normal distribution for all background rejections. These distributions therefore verify that the look elsewhere effect in this analysis is under control. 
In order to complete the frequentest study, the procedure must be repeated using many more statistically accurate samples. As the toy MC generator produces limited statistics, a Generative Adversarial Network (GAN) model must be used to generate statistically accurate samples at scale. The GAN model will therefore provide sufficient samples for the study, investigating the uncertainty generated in semi-supervised models, to be complete.